\documentclass[12pt]{article}
\usepackage{sint_modi,cite}
\usepackage{amsmath,amssymb}
\usepackage{epsfig}
\usepackage{amssymb}

\usepackage{amssymb}
\usepackage{epsfig}

\def\a{\alpha}
\def\b{\beta}
\def\g{\gamma}

\def\D{\Delta}
\def\d{\delta}
\def\e{\varepsilon}

\def\s{\sigma}

\def\beq{\begin{equation}}
\def\eeq{\end{equation}}
\def\beqn{\begin{eqnarray}}
\def\eeqn{\end{eqnarray}}
\def\ba{\begin{eqnarray}}
\def\ea{\end{eqnarray}}

\def\xprim2bar{\overline{x}^{\prime\prime}}

\def\beq{\begin{equation}}
\def\eeq{\end{equation}}

\newcommand{\beqa}{\begin{eqnarray}}
\newcommand{\eeqa}{\end{eqnarray}}

\let\a=\alpha   \let\b=\beta   \let\g=\gamma   \let\d=\delta
\let\e=\epsilon

\let\s=\sigma        

  \let\D=\Delta

\newcommand{\oV}{\overline V}

\newcommand{\ov}{\overline v}

\let\a=\alpha   \let\b=\beta   \let\g=\gamma   \let\d=\delta
\let\e=\epsilon

\let\s=\sigma        

  \let\D=\Delta

\newcommand{\be}{\begin{equation}}
\newcommand{\ee}{\end{equation}}
\newcommand{\bea}{\begin{eqnarray}}
\newcommand{\eea}{\end{eqnarray}}

\usepackage{graphics}
\usepackage{graphicx}
\newcommand{\eq}[1]{Eq.~(\ref{#1})}
\newcommand{\fig}[1]{Fig.~\ref{#1}}

\newcommand{\sect}[1]{Section~\ref{#1}}

\def\A5{(A_5)_{\rm lat}}
\def\thintablerule{\hrule height0.4pt}
\begin{document}

\rightline{WUB/09-15}

\vskip 1.5cm
\centerline{\LARGE A new model for confinement}

\vskip 2 cm
\centerline{\large Nikos Irges and Francesco Knechtli}
\vskip1ex
\centerline{\it Department of Physics, Bergische Universit\"at Wuppertal}
\centerline{\it Gaussstr. 20, D-42119 Wuppertal, Germany}
\centerline{\it e-mail: irges, knechtli@physik.uni-wuppertal.de}
\vskip 1.5 true cm
\thintablerule
\vskip 2.0ex
\leftline{\bf Abstract}

We propose a new approach towards the understanding of confinement.
Starting from an anisotropic five dimensional pure gauge theory, we approach 
a second order phase transition where the system reduces dimensionally.
Dimensional reduction is realized via localization of the gauge and scalar
degrees of freedom on four dimensional branes. 
The gauge coupling deriving from the brane Wilson loop observable 
runs like an asymptotically free coupling at
short distance, while it exhibits clear signs of string formation at long
distance. The regularization used is the lattice. We take the
continuum limit by keeping the ratio of the lattice spacing in the brane over
the lattice spacing along the extra dimension constant and smaller than one.

\vskip 1.0ex\noindent
\vskip 2.0ex
\thintablerule

\vskip-0.2cm
\newpage

\section{Introduction}

Four dimensional $SU(N)$ gauge theories have a unique phase in which interactions are confined
\cite{Confinement}.
The only known fixed point in their phase diagram which is at vanishing gauge coupling, $g^2=0$,
is an ultraviolet (UV) fixed point where weak coupling perturbation theory is a good description. 
According to perturbation theory the interactions in this regime are
dominantly 
of Coulomb type, with the charge obeying
the renormalization group (RG) flow appropriate to an asymptotically 
free coupling\cite{AsymptFreed}.
Perturbation theory is however oblivious to the long distance effects of confinement.
In order to see those, one must increase the coupling to larger values where
the only probe we know of are lattice \cite{Wilson} Monte Carlo (MC) simulations.\footnote{At very strong 
coupling we can also use strong coupling expansion methods but the information
we can extract from them is limited and in addition there is no 
analytic connection to the weak coupling regime.}
As the MC simulations reveal, globally the static potential seems to consist of two distinct regimes. 
At short distance it is indeed of a Coulomb form 
but at long distance it develops a linearly growing behavior. 
More precisely, at short distance the
coupling $\a=g^2/4\pi$ defined as (for $SU(2)$)
\be
\alpha_{q\bar{q}}(1/r) = -\frac{4}{3}c(r)\,, 
\ee
with $c(r)=(1/2)r^3F'(r)$ and $F$ the static force,
decreases according to the perturbative RG flow
and around a certain scale $r_s$,
defined as the scale where perturbation theory breaks down,
$c(r)$ plateaus around the value $-0.3$ \cite{LuescheWeisz}. 
The physically motivated explanation of this behavior comes 
from an effective string description according to which confinement results 
in the formation of a string like flux tube.
The positive slope of the linear term in the static potential is then 
interpreted as the tension $\s$ of this string.
The massless degrees of freedom that describe the 
fluctuation of the tube are Goldstone modes with an effective action that
can be written in the form of a derivative expansion constrained by Poincar{\'e}
(and perhaps also diffeomorphism) invariance. 
The kinetic term in this class of "world-sheet" actions, when integrated out, yields a universal
term, the L\"uscher term \cite{Luscherterm}, with value $-(d-2)\pi/(24r)$ in
$d$-dimensions.
This is the term that the MC sees as a plateau in $c(r)$ at large distance.
Similar properties are believed to hold for any generalization of the pure 
gauge theory where the string is stable \cite{Kuti}.
In summary, the description of the static potential from short to long distance entails
two different analytic methods, namely
weak coupling perturbation theory and an effective string description, each of which is
blind to the physical effects that the other sees. The two approaches are bridged by
the MC which, in principle, can probe the whole static potential from weak to strong coupling.
In practice however \cite{Necco:2001xg}, at short distance, simulations tend
to give a much less precise description compared to 
the usual continuum field theory Feynman diagrams due to the enhanced lattice artifacts and 
the same applies to any analytic perturbative lattice computation. 
It is fair to say that a global analytic understanding of the static potential
of confined 4d gauge theories is missing. 

In \cite{Irges:2009bi} we proposed a regularization of four dimensional gauge theories which could allow
for such a unified description. 
The idea is to start form an $SU(N)$ pure gauge theory in 5d and via a systematic  
expansion around an anisotropic mean-field background \cite{Meanfield} try to
reach an ultraviolet fixed point
(or points) in the interior of the phase diagram where a second order phase transition takes place.
The anisotropic background has the value $\ov_{05}$  along the extra dimension and 
$\ov_0$ along the other four directions (see Appendix A).
In the confined phase it vanishes while in the deconfined phase it is non-zero along all directions.
There is a (unstable according to the leading order mean-field method) 
phase where $\ov_0\ne 0$ and $\ov_{05}=0$, the layered phase.
Only on the isotropic lattice the background is isotropic.
A line of second order phase transitions was found on the boundary between the deconfined phase
and the layered phase. From the side of the deconfined phase, near the phase transition,
surprisingly, the system reduces dimensionally. 
Even though not the same, the physics in this phase is similar in spirit 
to the layered phase of \cite{Layered}.
We have called in \cite{Irges:2009bi} this phase the "d-compact phase".
The low energy degrees of freedom of the dimensionally reduced system are 
those of the four dimensional Georgi--Glashow model
which is in the class of theories described in the first paragraph.
The technical tool used to carry out the expansion is the lattice regularization.
This allows for a well defined description and control of the quantum theory but there is a price.
Because the final results for the observables are expressed in terms of finite lattice 
sums it is not easy to take the continuum limit analytically. 
We emphasize though that this is merely a technical obstruction.
The method being fully analytic should, in principle, allow one to carry out
all limits without resorting to numerical methods.
This will be attempted at a later stage.
Here we perform a numerical analysis of the analytic results of
\cite{Irges:2009bi}, with our main focus
on the static potential. We will take carefully the continuum limit and 
we will try to argue that the static 
potential oriented along the four dimensional hyperplanes, 
computed in this scheme, reflects both the asymptotically free and the confining aspects of the coupling.

The mean-field expansion comes with certain caveats. 
The mean-field background is gauge dependent. Within the class of Lorentz
gauges we found our physical observables independent of the gauge fixing
parameter $\xi$ to leading order \cite{Irges:2009bi}.
There is no guarantee that the expansion converges. It is known however that the corrections come 
multiplied by powers of $1/d$ and therefore convergence is expected to 
become better as the number of dimensions $d$
increases. The mean-field sometimes fakes phase transitions. 
Even though the known such cases are generally less sophisticated 
compared to our construction, it is still   
conceivable that there is an intricate way in which the mean-field does generate a fake picture.
This is where the MC investigation of the phase diagram of these theories will
be crucial \cite{MC}.
The reason we proceed with presenting the results of the mean-field ignoring this possibility is that
even though the MC may not see an UV fixed point where the system reduces dimensionally,
the approach seems to have an independent value. 
It can serve as a new analytic laboratory of confining gauge theories from short to long distance. 

\section{The model and its Lines of Constant Physics}

The model considered in \cite{Irges:2009bi} is an anisotropic $SU(2)$ 
lattice gauge theory in 5 dimensions
defined in a mean-field background. 
Gauge theories in five dimensions are defined on an anisotropic, infinite, hypercubic lattice via two
independent dimensionless parameters. One is $\beta$, the lattice coupling and
the other is $\gamma$, the anisotropy parameter.
One way to define them is through the dimensionful parameters of the lattice. 
We consider first finite hypercubic lattices with the same number of lattice points
along the four dimensions $L=l / a_4$ and along the fifth dimension $L=2\pi
R /a_5$ and eventually take the $L\to \infty$ limit. $l$ and $R$ are the 
physical sizes and as appropriate to anisotropic
lattices we take different lattice spacings along the four and fifth
dimensions. The coupling of a five dimensional gauge theory is
denoted by $g_5$ and it has dimension of $\sqrt{\rm length}$.
The anisotropy parameter can be defined at 0$th$ order as
$\gamma = a_4/a_5 = l/(2\pi R)$  
and the $SU(2)$ coupling as $\beta = 4a_4/g_5^2$.
In these variables, the perturbative regime is located at $\beta
= \infty$. The above mentioned d-compact phase appears instead for $\beta\sim O(1)$ and
$\gamma <1$, obviously far from perturbation theory.
The line of second order phase transition extends in a range 
that corresponds to approximately $\g < 0.62$. 
Physically, this is a situation where the extra dimension is
larger than the spatial directions and gauge interactions are 
localized on four dimensional hyperplanes.

The action used to compute observables was the Wilson plaquette action
\be
S_W =
\frac{\b}{4} \Bigl[ \frac{1}{\g}\sum_{\rm 4d-p} \Bigl(1-{\rm tr}\{U_p \} \Bigr)
+ {\g}  \sum_{\rm 5d-p} \Bigl(1-{\rm tr}\{U_p \}\Bigr) \Bigr],\label{anisaction}
\ee
where the first term contains the effect of all plaquettes $U_p$
along the four dimensional slices of the five dimensional space and the second term contains the effect 
of plaquettes having two of their sides along the extra dimension.
In the mean-field, for $SU(2)$, the fluctuating degrees of freedom are complex 
valued quantities $V(n,M)$ located at the lattice site $n$ and pointing in the direction $M$.
The schematic expression for the correction to the expectation value of a physical,
gauge invariant observable ${\cal O}$ to second order in the mean-field expansion has the form 
\bea
<{\cal O}> &=& {\cal O}[\overline V]+  \frac{1}{2}\left(\frac{\d^2  {\cal O}}{\d
  V^2}\right)_{ik}\left(K^{-1} \right)_{ik} \nonumber\\
&+&\frac{1}{24}\sum_{i,k,l,m}\left(\frac{\d^4  {\cal O}}{\d
  V^4}\right)_{iklm}\Bigl( (K^{-1})_{ik}
(K^{-1})_{lm}+(K^{-1})_{il}(K^{-1})_{km}+(K^{-1})_{im}(K^{-1})_{kl}\Bigr)\nonumber\\
\label{2correction}
\eea
where $K$ is the appropriate lattice propagator and the sums over $i,k,l,m$ are sums over links. 
Derivatives and contractions are taken in the mean-field background.
All such quantities are gauge independent. 
Polyakov loops with scalar and vector transformation properties represent corresponding 
classes of states in the Hilbert space and from the exponential decay of their
Euclidean time correlators their mass spectra can be extracted.
The expectation value of Wilson loops can be used to extract the static potential.
The anisotropic lattice admits two inequivalent classes of Wilson loops.
One class consists of the loops oriented along the time direction 
and one of the spatial directions. The other class consists of loops along
the time and the extra dimension, 
the latter defined as the direction along which the background is different from the 
background along the other four directions.  
We call the static potentials derived from these two classes as $V_4$ and $V_5$ respectively
and define as $F_j(r-a_j/2) = \{V_j(r)-V_j(r-a_j)\}/a_j$, $j=4,5$ the corresponding forces. 

The dimensionless vector mass (see \eq{W})
is found to have a dependence only on the lattice size\cite{Irges:2009bi}
\be
a_4m_W  =  c_L/L \,,
\ee
where $c_L$ is a constant with numerical value $12.61$.
Therefore the system can not be in the Higgs phase in a mean-field background.
Hence, if dimensionally reduced, it must be in a confined phase.
The scalar mass $a_4m_H$ (see \eq{H}) can be used to measure a critical 
exponent of $\nu=1/2$ as the second order phase transition is approached. 
The potential $V_5(r)$ (see \eq{V5}) determines the interaction of static quarks along the extra dimension.
The static potential $V_4(r)$ (see \eq{V4}) is the quantity that dictates the behavior of the 
dimensionally reduced coupling and it is well defined
for any value of the distance $r$. For more details on these computations 
we refer the reader to \cite{Irges:2009bi}.
 
It is convenient to trade the bare lattice coupling $\b$ for the
dimensionless physical ratio $\rho=a_4m_W/a_4m_H$ and parametrize the model
by the two dimensionless physical ratios $\rho$ and $\gamma$.
The lattice spacing can be then chosen to be measured by $a_4m_H$. 
An issue arises when one realizes that on an infinite lattice 
both the numerator and the denominator of $\rho$ approach zero in the limit $a_4\to 0$. 
This means that the continuum limit must be reached in a way that regularizes $\rho$. 
The method to achieve this is to approach the phase transition
at fixed $\gamma$ and at every step, adjust $L$ and $\beta$ so that $\rho$ remains constant. 
When taken in this way, the infinite lattice and continuum limits coincide:
\be
(L\longrightarrow \infty,\;\; \b\longrightarrow \b_c)|_{\g, \rho = {\rm const}}\;\; 
\Longleftrightarrow \;\; {\rm continuum \; limit}\label{LCP}
\ee
A crucial fact that allows one to apply this method without obstacles
is the fact that $a_4m_W$ depends only on $L$ and that $a_4m_H$ depends only on $\beta$ and $\gamma$.
We call from now on such continuum limit trajectories on the phase diagram 
approaching the phase transition Lines of Constant Physics (LCPs).
The whole discussion above can be generalized to any $SU(N)$ gauge group. 
Here we concentrate only on $SU(2)$.

\section{The continuum limit}

Following \cite{Sommer:1993ce}
we define a physical scale through the condition
\bea
r_{\rm s}^2F_4(r_{\rm s}) = s = 0.2 \,.\label{sommer}
\eea
$s=0.2$ is chosen so that $r_s/a_4$ lies in the transition region from the short
distance behavior of the force to the long distance one. 
%
\begin{figure}[!t]
\begin{minipage}{8cm}
\centerline{\epsfig{file=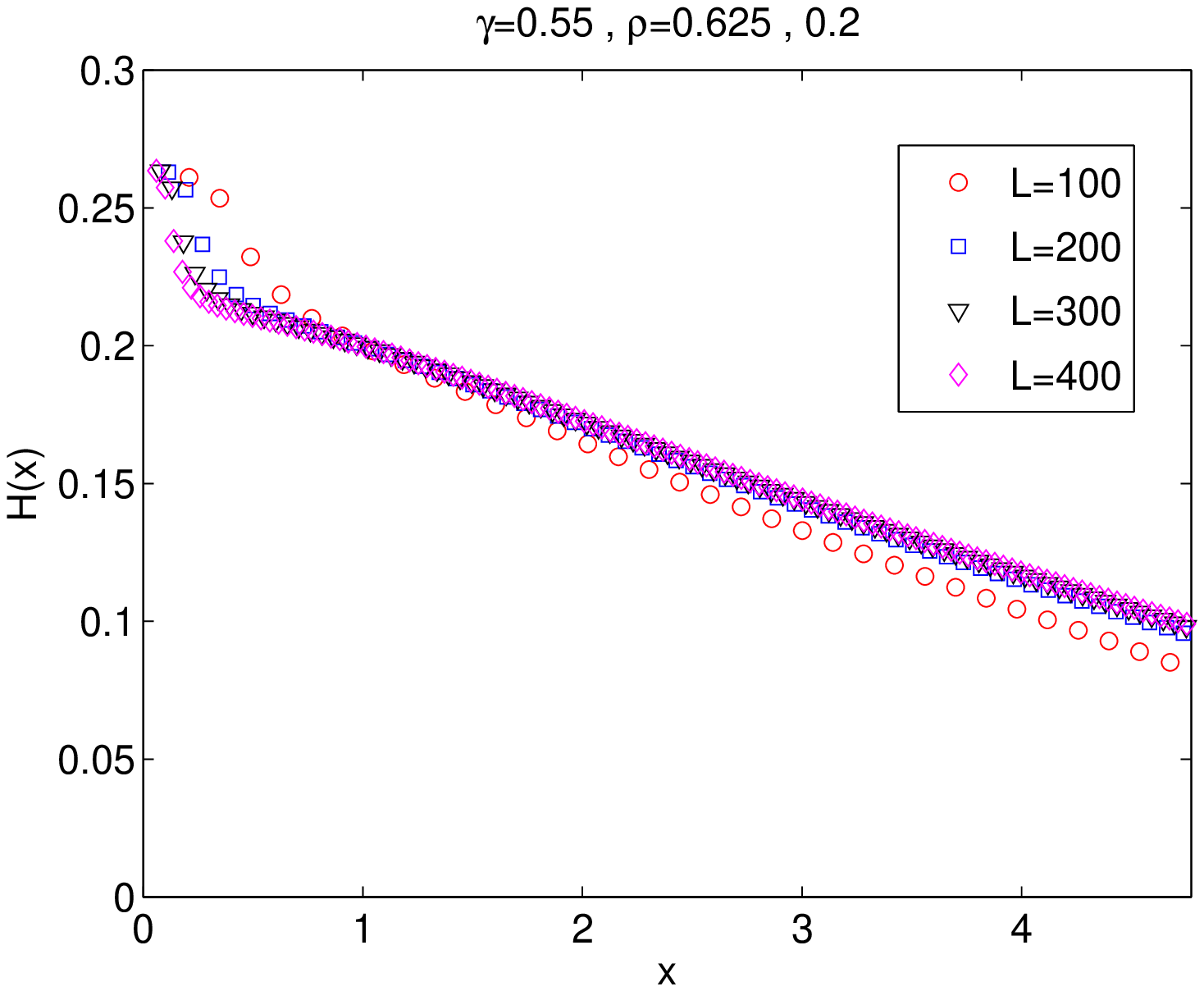,width=8cm}}
\end{minipage}
\begin{minipage}{8cm}
\centerline{\epsfig{file=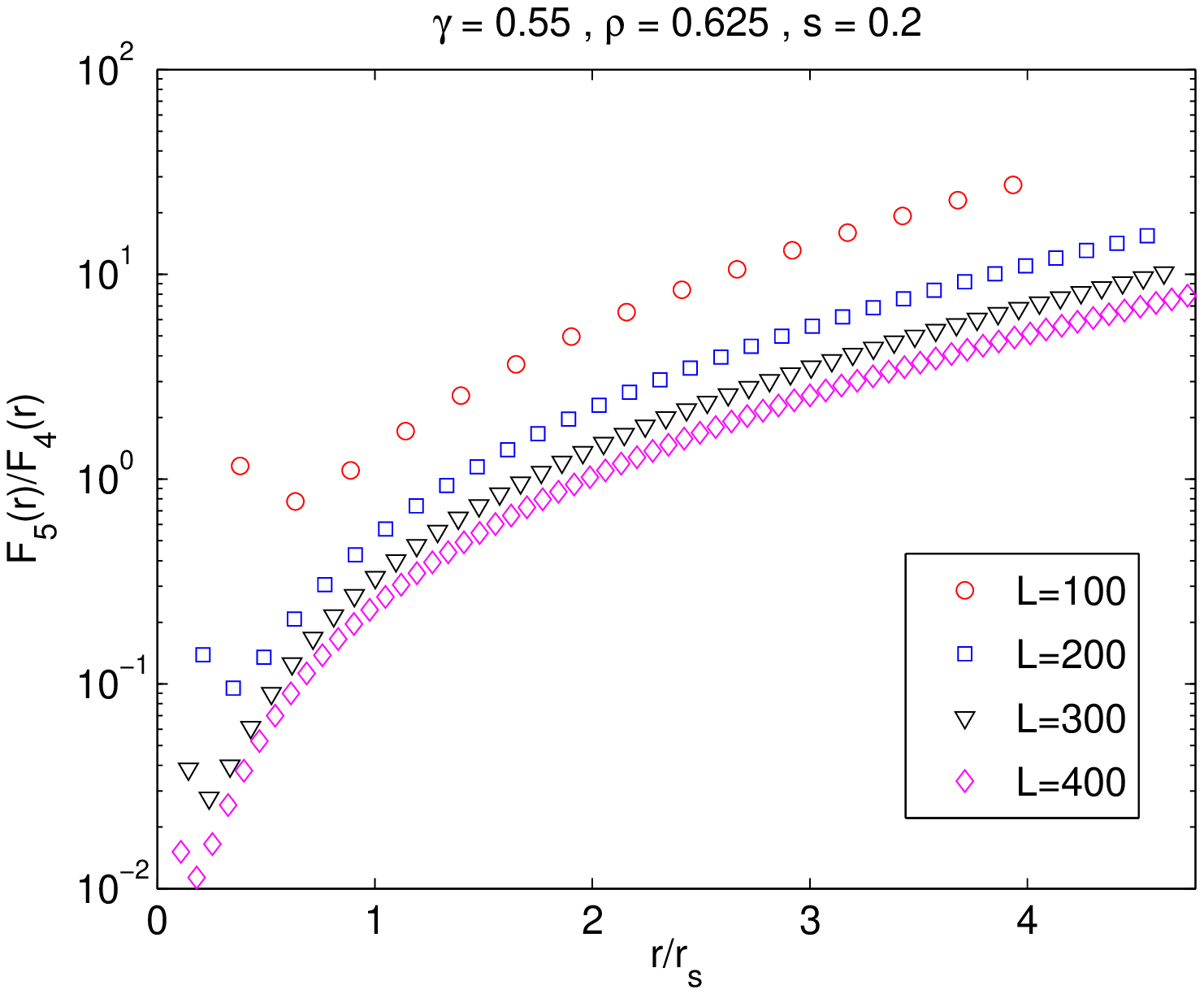,
width=8cm}}
\end{minipage}
\caption{\small Left Plot: the function $H(x)$ in \eq{Hdef} along the LCP A. 
Right plot: the ratio of the extra dimensional to four dimensional force
$F_5(r)/F_4(r)$ along the LCP A.
\label{f_force_A}}
\end{figure}
%
We will first consider the LCP trajectory defined as
\be
{\rm LCP\; A}:\;\;\; \g=0.55\;\; {\rm and}\;\; \rho=0.625 \,,
\ee
then consider other values of $\rho$ for the same $\g$ and finally 
we will analyze other LCPs labelled by values in the range $0.62 > \g > 0$
(i.e. move down along the line of second order phase transitions). 

In the left plot of \fig{f_force_A} we show the dimensionless function
\be
H(x) = \left. F_4(r)r^2\right|_{r=xr_s} \label{Hdef}
\ee
for various lattice spacings corresponding to $L$ values in the range
$100,\ldots,400$. The function shows good scaling. The shape of $H(x)$
is different than in a pure $SU(2)$ gauge theory, see \cite{Sommer:1993ce},
in particular it decreases as $x$ increases. We will see below that this is
due to the presence of a positive $1/r^2$ contribution at short disctance and
a large negative logarithmic contribution at large distance 
in the static potential $V_4$.

In the right plot of \fig{f_force_A} we show the ratio $F_5(r)/F_4(r)$ on a logarithmic scale as
a function of $r/r_s$ for different lattices along the LCP A. When we take the
continuum limit the ratio decreases steadily which implies that the force $F_5$ is not physical.
This is an evidence for localization: since $F_4$ has a finite continuum limit
(see left plot in \fig{f_force_A}),
$F_5$ tends to zero, implying that the five dimensional space
decomposes into a set of non-interacting Euclidean four-branes along the fifth dimension.
On each of these branes the localized light degrees of freedom are therefore 
expected to be those of the four dimensional Georgi--Glashow model. 
Since the mean-field background does not break the gauge symmetry spontaneously,
these degrees of freedom must be in the confined phase.
In the following, we will take the continuum limit of the static potential $V_4$ and try to 
argue that our picture of dimensional reduction is consistent.

To extract physical information from the static potential \eq{V4} we start from an ansatz of the form
\be
V(r) = \mu + \sum_p \frac{c_p}{r^p} \,,
\ee
with $p\ge -1$ ($p=0$ corresponds to a logarithmic term and $c_{-1}\equiv \sigma$), solve locally
for the coefficients $c_p$ and plot them as a function of $r$ 
(in practice we will rescale all dimensionful quantities by
appropriate powers of $r_s$ to make them dimensionless). 
The limit $L\to \infty$ will be taken by computing the coefficients for increasing values of 
$L$ and extrapolating to the infinite lattice.

\subsection{Short distance \label{ss_short}}

For $r/r_s<1$ we choose the ansatz
\be
V(r) = \mu + \frac{c_1}{r} + \frac{c_2}{r^2} \,, \label{Short}
\ee
the role of the $1/r^2$ term being the check of an imperfect dimensional reduction. 
%
\begin{figure}[!t]
\begin{minipage}{8cm}
\centerline{\epsfig{file=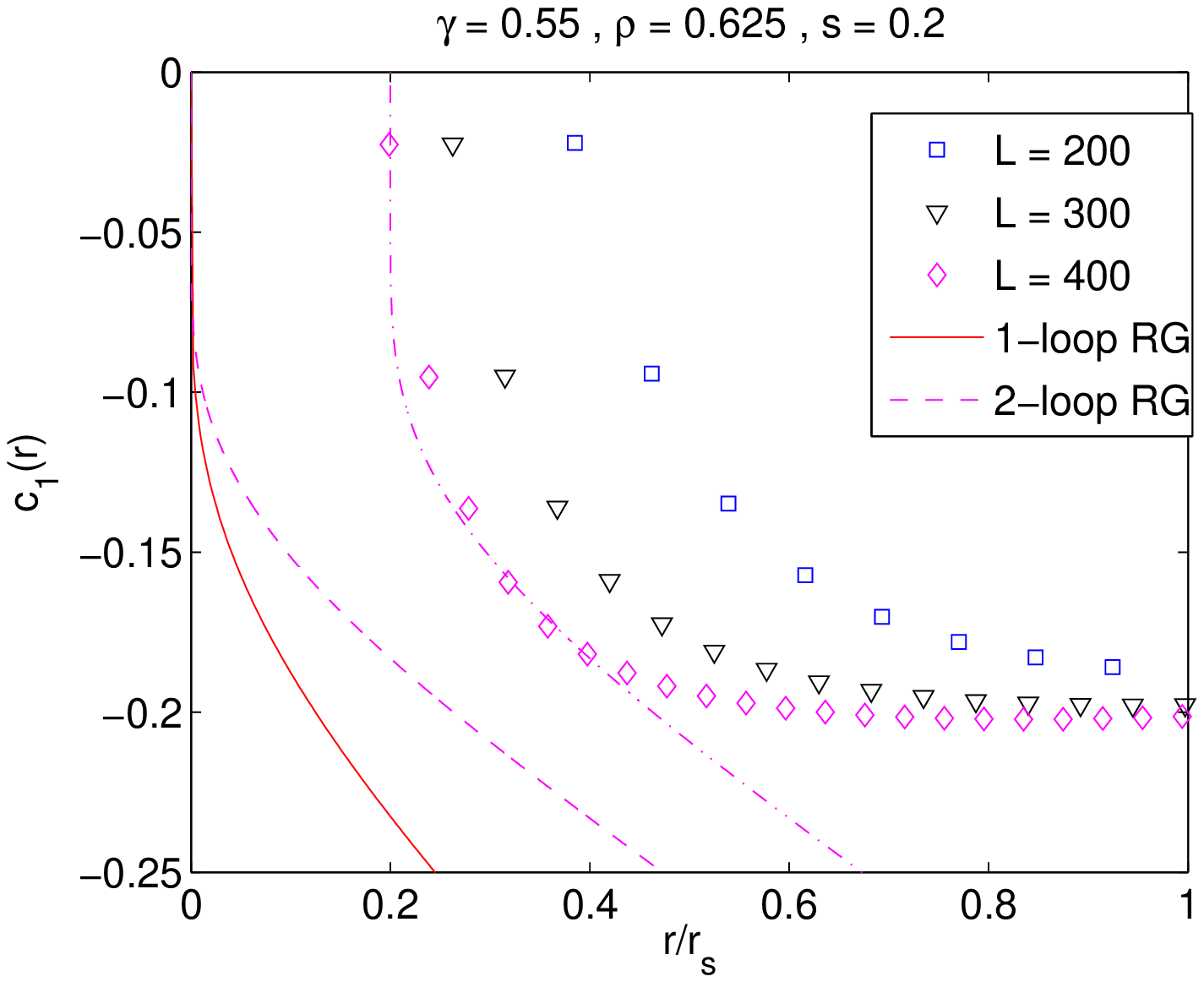,width=8cm}}
\end{minipage}
\begin{minipage}{8cm}
\centerline{\epsfig{file=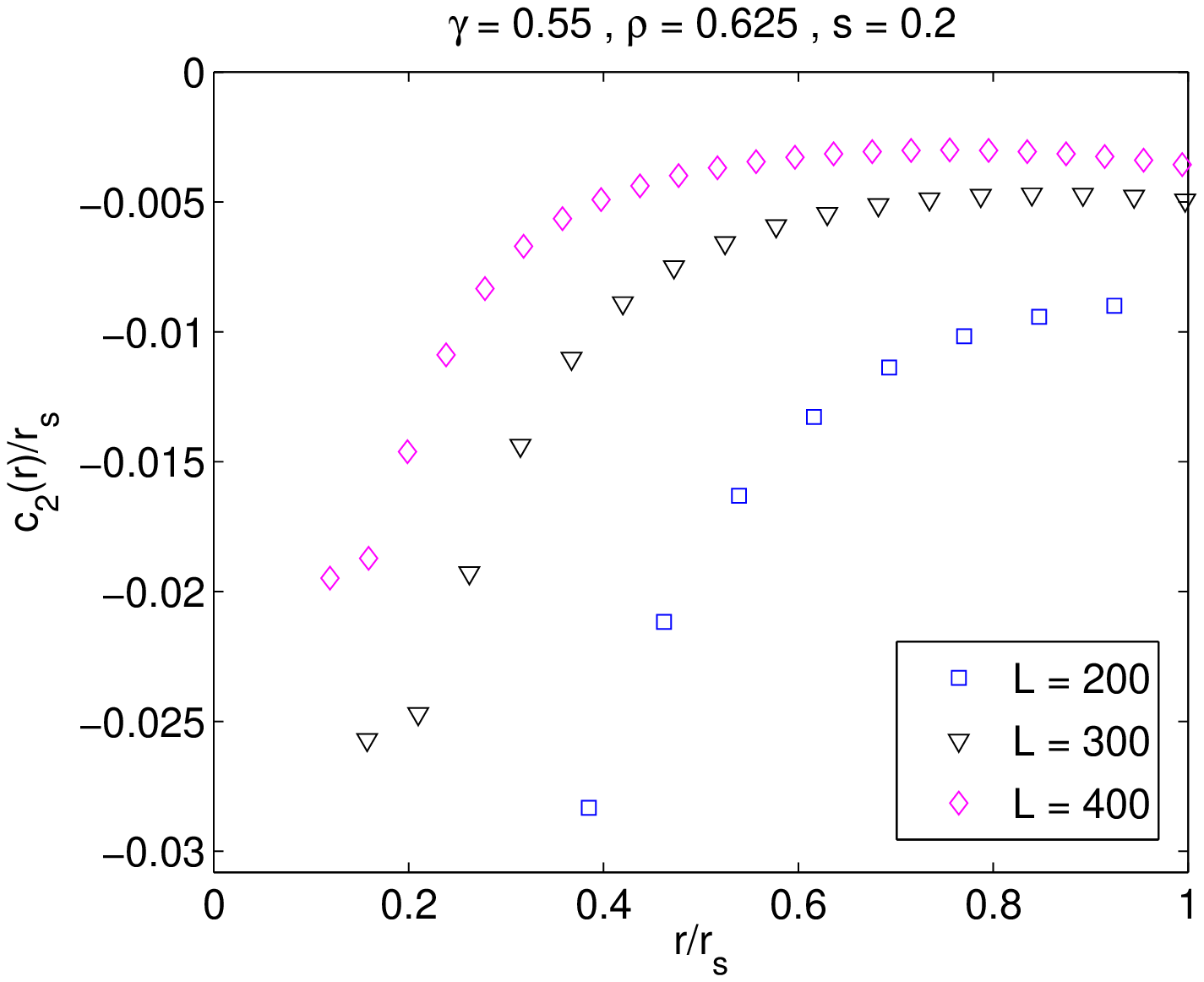,width=8cm}}
\end{minipage}
\caption{\small The short distance coefficients $c_p$ for the LCP A.
\label{shortRG}}
\end{figure}
%
In \fig{shortRG} we compute $c_1$ and $c_2$ (see Appendix B) and
compare the coefficient $c_1$ to the 1-loop and 2-loop RG evolution formula
\be
\Lambda_{q\bar{q}}\, r = \left(b_0 g^2 \right)^{-b_1/(2b_0^2)} e^{-1/(2b_0g^2)} \,,
\label{Lambda2l}
\ee
where $g^2=-(16\pi/3)\, c_1$ and $g^2=g^2(1/r)$. 
For the Georgi--Glashow model $b_0=7/(4\pi)^2$ and $b_1=(115/3)/(4\pi)^4$.
The 1-loop formula is obtained for $b_1=0$.
Unlike for QCD there is no available  experimental or lattice 
MC prediction for a scale, so we will choose it according to
convenience to $\Lambda_{q\bar{q}}\, r_s=0.277$. 
Also, due to the high degree of the discrete derivatives involved in the determination of $c_1(r)$, 
the latter can be defined only for $r/a_4>2$ which makes the comparison with the RG formula
valid near $r=0$ hard. Nevertheless, it is clear from the left of
\fig{shortRG} that even though the continuum limit has not been reached, as $L$ increases there is a 
stable convergence to the theoretical curve. 
The continuous line on the left is the 1-loop curve and the dashed one on the right is the 2-loop curve.
To illustrate our point we have shifted the 2-loop curve by a constant, shown
as a dashed dotted line.   
In addition, according to the right
of \fig{shortRG}, as $L\to \infty$, the $1/r^2$ piece gradually disappears.

\subsection{Long distance}

\begin{figure}[!t]
\begin{minipage}{8cm}
\centerline{\epsfig{file=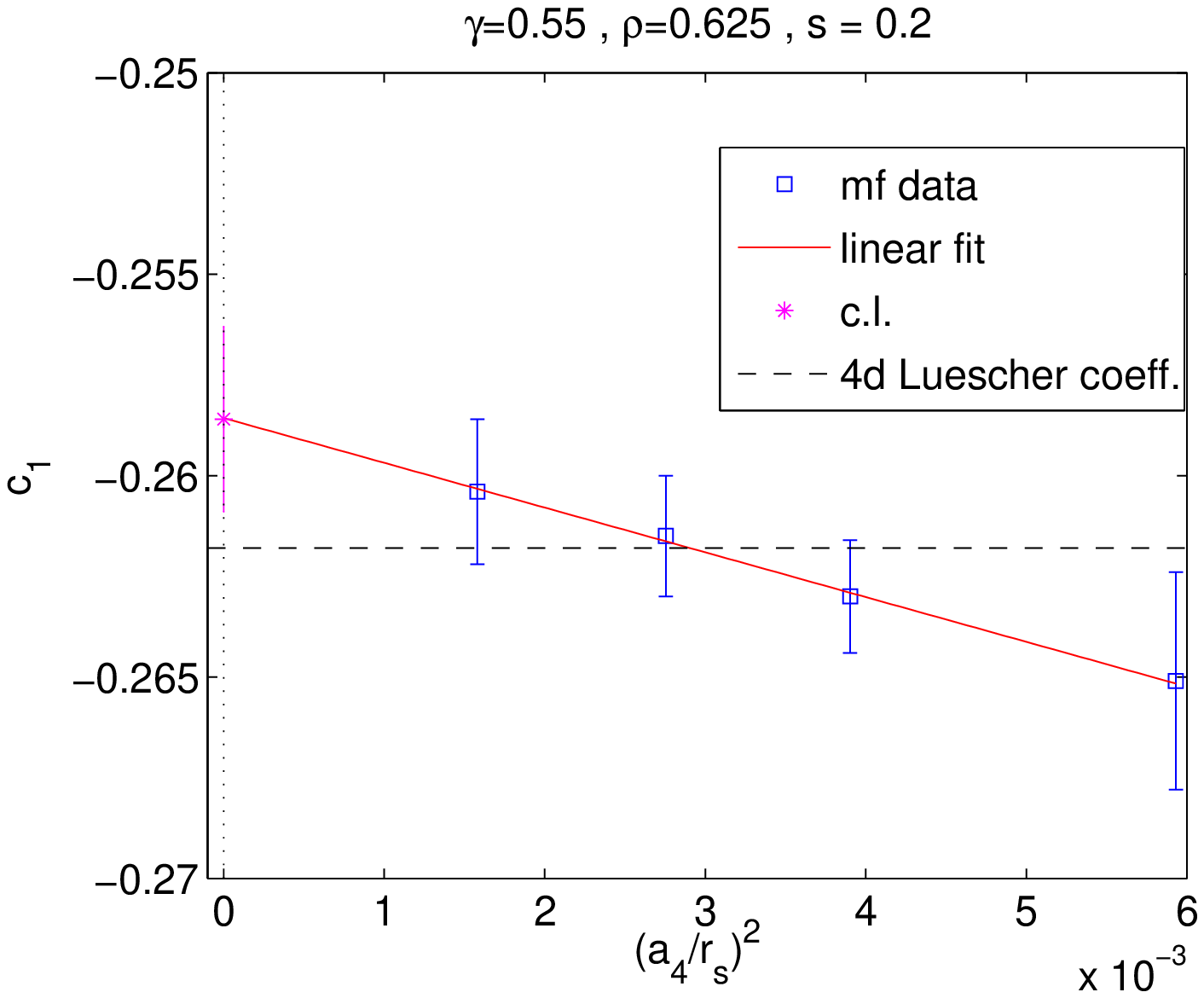,width=8cm}}
\end{minipage}
\begin{minipage}{8cm}
\centerline{\epsfig{file=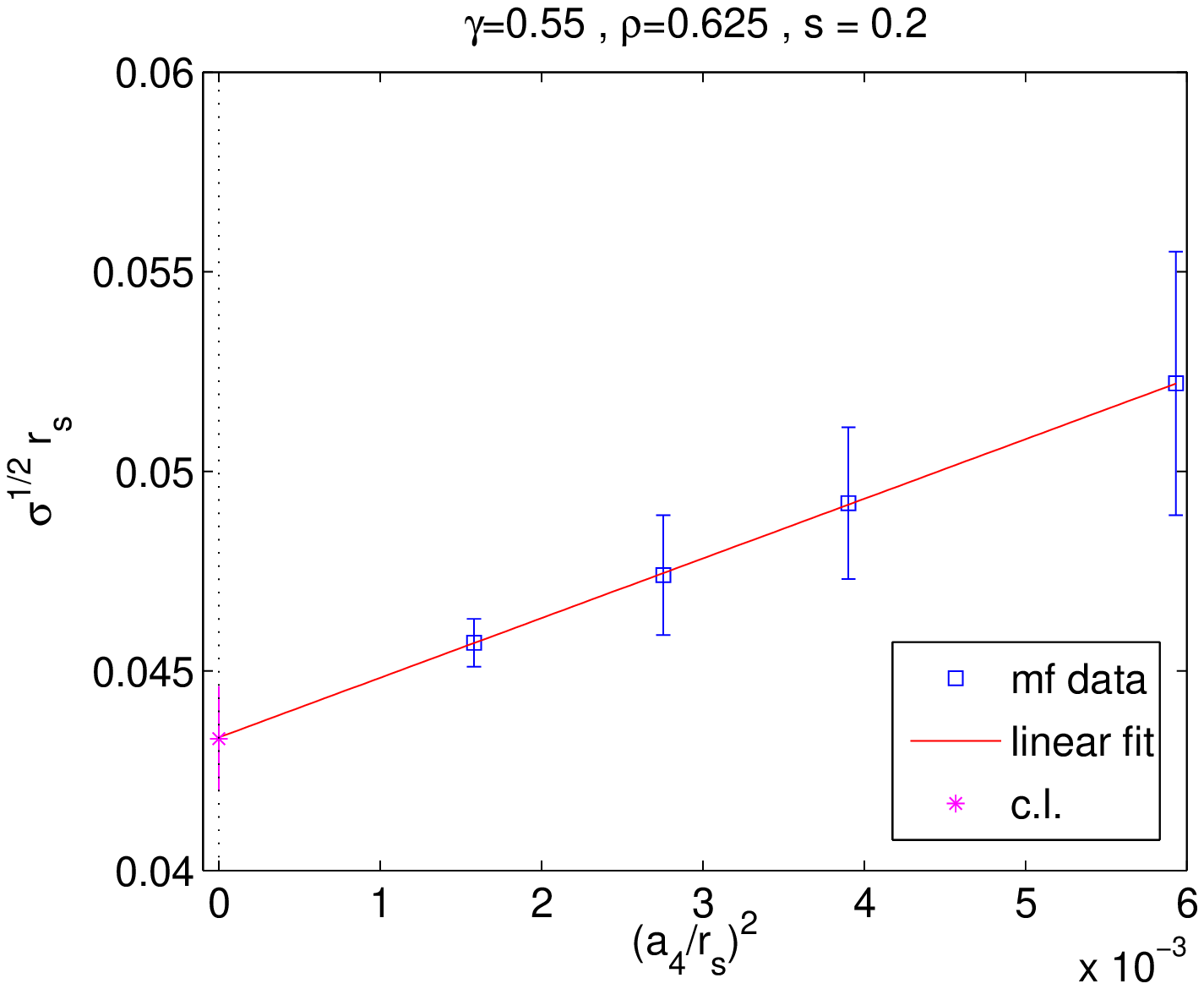,width=8cm}}
\end{minipage}
\begin{minipage}{8cm}
\centerline{\epsfig{file=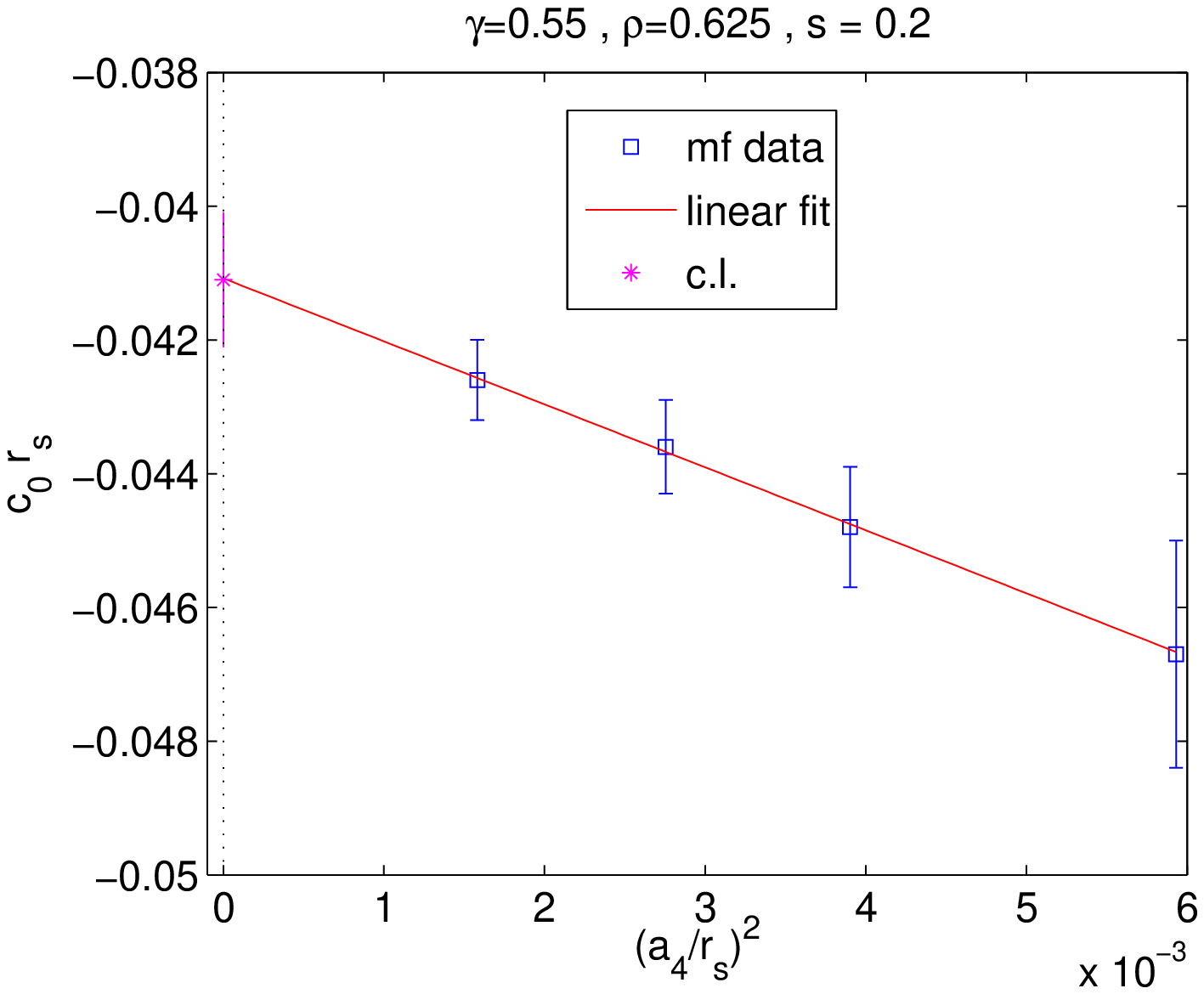,width=8cm}}
\end{minipage}
\begin{minipage}{8cm}
\centerline{\epsfig{file=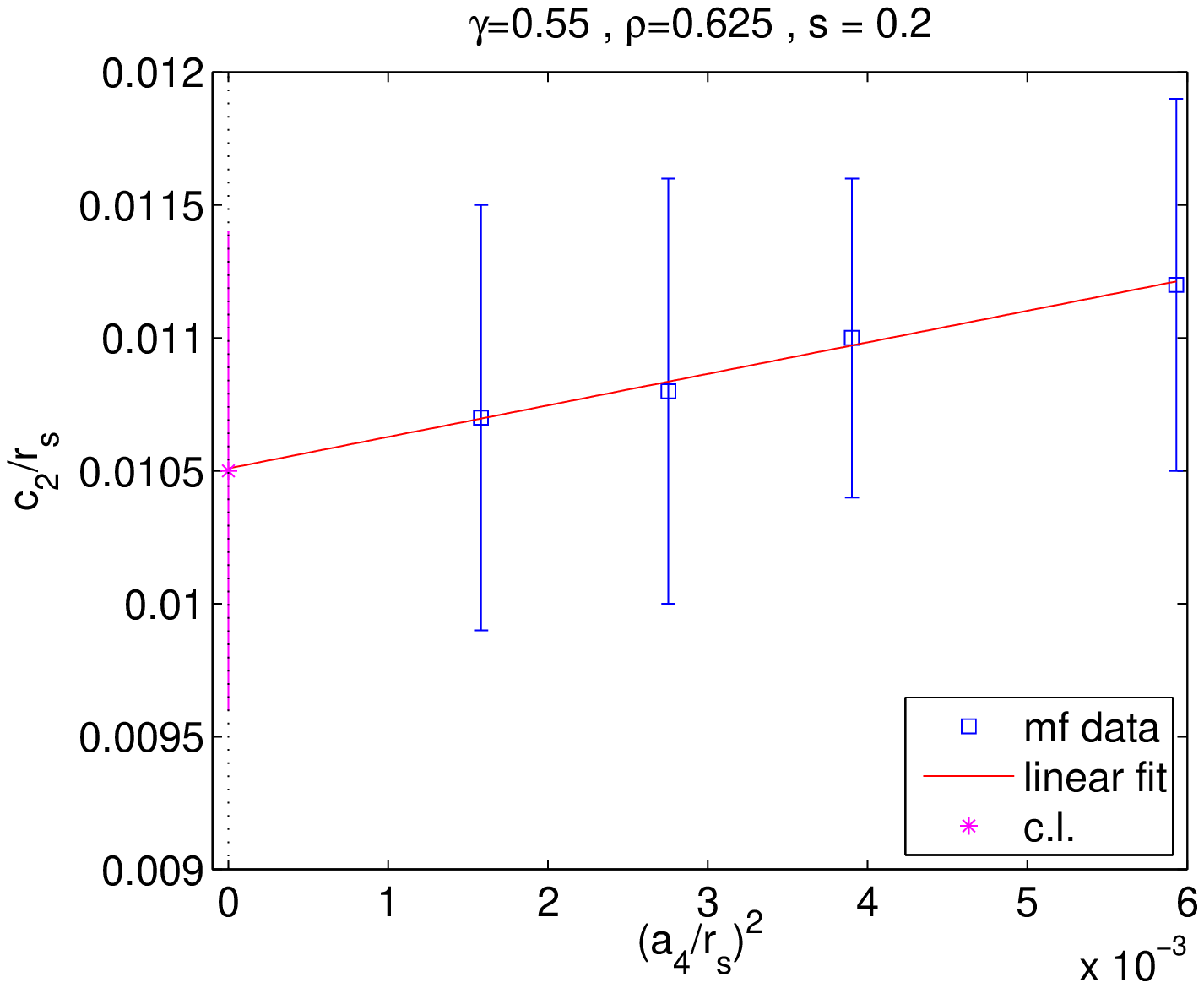,width=8cm}}
\end{minipage}
\caption{\small The continuum limits of the long distance coefficients $c_p$ for the LCP A.
\label{LCPA}}
\end{figure}

\begin{figure}[!t]
\begin{minipage}{8cm}
\centerline{\epsfig{file=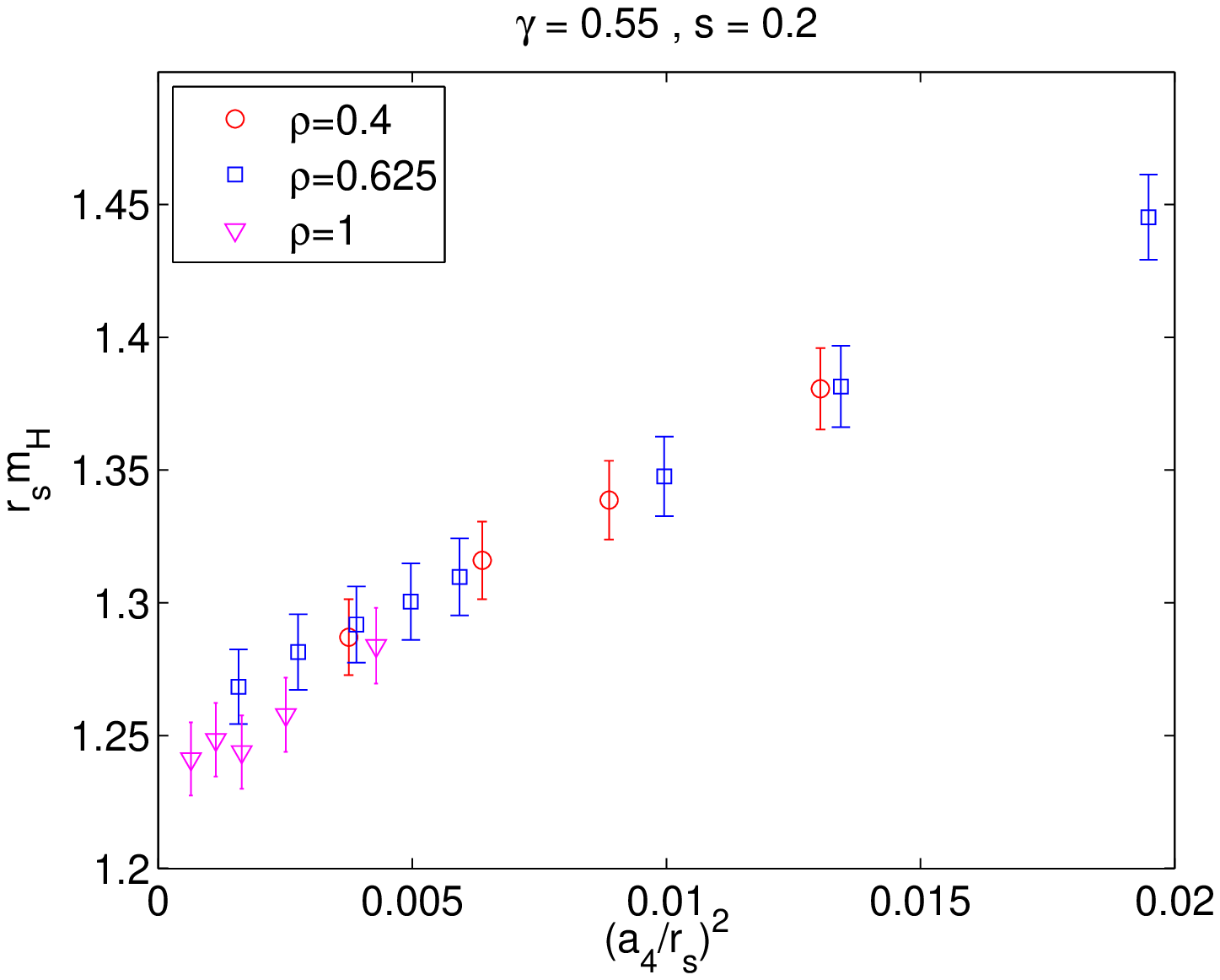,width=8cm}}
\end{minipage}
\begin{minipage}{8cm}
\centerline{\epsfig{file=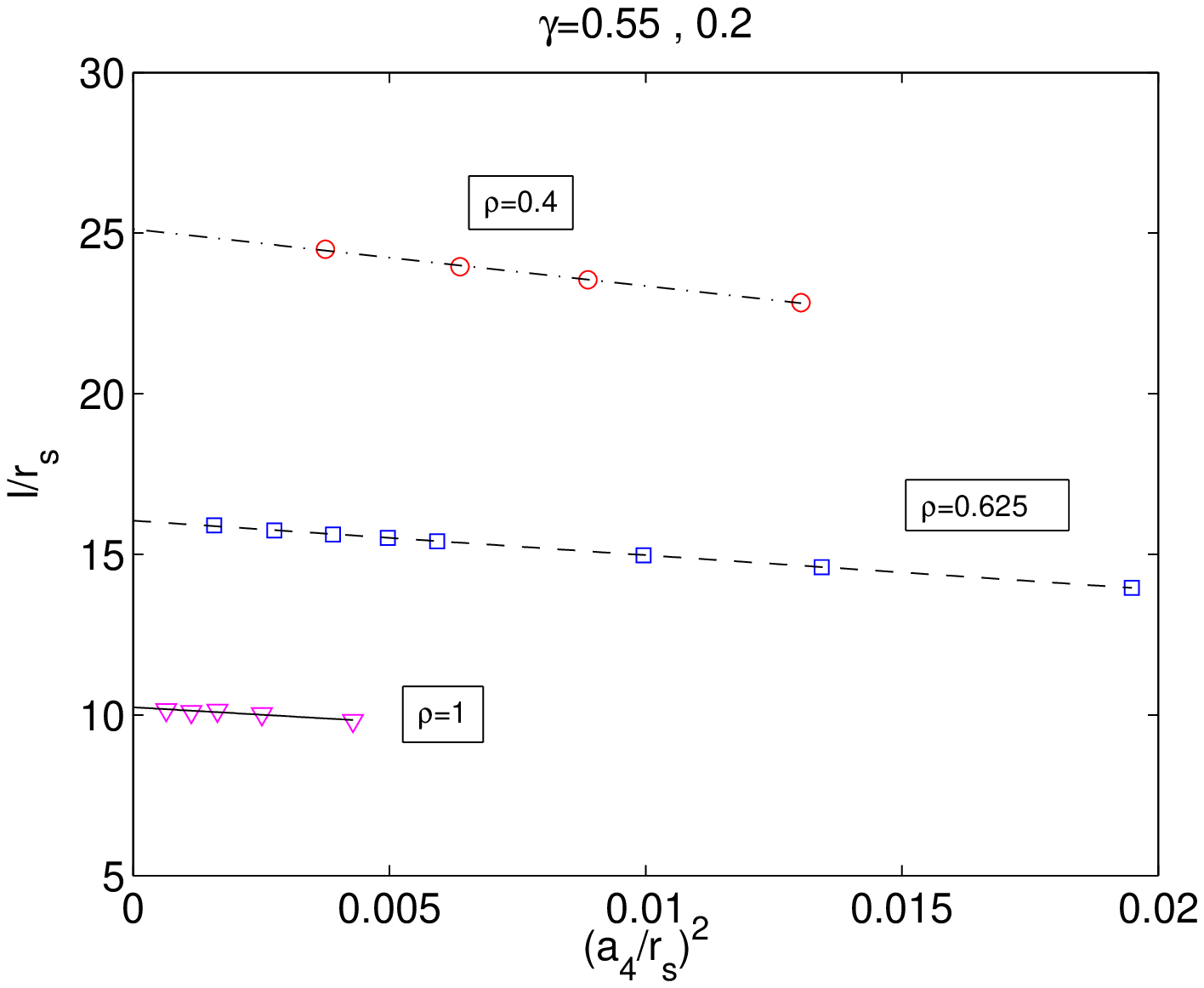,width=8cm}}
\end{minipage}
\caption{\small The continuum limits of $r_s m_H$ and $l/r_s$. 
The errors of $m_H$ originate from the ones of $m_W$ and come from the 
uncertainty in the effective mass at $L=100$. 
$m_W$ is scaled proportional to $1/L$. On the right linear fits in $a_4^2$
show the continuum limit.
\label{rhovar1}}
\end{figure}

For $r/r_s>1$ we choose the ansatz
\be
V(r) = \mu + \s r + c_0 \log(r) + \frac{c_1}{r} + \frac{c_2}{r^2}
\label{Long}
\ee
and compute the coefficients using the discretized formulae in Appendix B.
The plots show plateaus forming for all four coefficients at the same range of distances
\be
r/r_s \in [2.15\, , \, 2.80], \label{plateau}
\ee
a sign that the ansatz is close to the actual form.
Moreover we have checked that the plateaus form essentially independently from the way we
discretize the derivatives of $V(r)$.
We take as the plateau value of a coefficient the average of the quantity over the range 
indicated in \eq{plateau}. 
Reading off the value of a coefficient (or of a mass) from a plateau introduces errors, 
which we add on its plot. 
In \fig{LCPA} we compute the continuum limit of the $c_p$ for the LCP A using
a linear fit in $a_4^2$, which is the expected form of leading cut-off effects
in the static potential \cite{Necco:2001xg}.
%
%
The continuum limit value of the string tension is
\be
\sqrt{\sigma}\, r_s = 0.0433(13)
\ee
and using the scale $\Lambda_{q\bar{q}}$ estimated in \sect{ss_short} we obtain roughly
$\Lambda_{q\bar{q}}/\sqrt{\sigma} \simeq 6.4$. 
Even though not from the same theory, the pure four dimensional $SU(2)$ gauge
theory value \cite{Luscher:1992zx,Sommer:1993ce}
(which is the only quantitatively reliable case we know) 
$\Lambda_{q {\bar q}}^{({\rm YM})}/\sqrt{\sigma} \simeq 0.68(10)$ tells us
that the string tension is probably rather small in our model.
A physical understanding of this fact can be obtained by 
reading off the continuum limit value 
\be
\sqrt{\sigma}\, l = 0.69
\ee
from \fig{LCPA} and \fig{rhovar1}:
the box seems to be barely sufficiently large to sustain a stable string.
In order to increase this number, one would need to increase $l$, which as we will
argue in the following, requires decreasing $\rho$ and increasing $L$. 
On the other hand, the continuum limit value of the coefficient $c_1$ is  
\be
c_1 = -0.2586(23)\, ,
\ee
in excellent agreement with the universal value of the $d=4$ L\"uscher
coefficient $-\pi/12$, as it should. 
The somewhat surprising term that seems to be though necessary for a correct interpretation
of our formulae is the $\log(r)$ term. In fact, if left out, a consistent picture can
be hardly obtained.
Regarding the $1/r^2$ term that we seem to be also seeing, as well as 
other possible higher negative powers 
of $r$ we would not like to make any committing statements until a more complete understanding of the 
structure of the lattice formula \eq{V4} both from the field theory and 
the effective string point of view is obtained.
%
%

The next issue concerns the universality of the continuum limit.
It is easier to formulate the question in
physical terms. In \fig{rhovar1} we compute the continuum limits of $r_s m_H$ (left)
and of $l/r_s$ (right) for three RG trajectories at $\gamma=0.55$ with different
$\rho$ values of $0.4$, $0.625$ and $1$. 
Evidently, $r_s m_H$ has (within errors) a $\rho$ independent continuum limt, around $1.25$.
$\rho$ on the other hand, when $L$ is varied, controls the physical size of the
lattice, i.e. the size of the box: as $\rho$ increases, the box shrinks.
When $L$ is kept constant, it controls the
lattice spacing: as $\rho$ increases, the lattice spacing decreases.  
It is interesting to notice that
since $\rho\, (l/r_s)$ has a universal continuum limit which is around $10$, and
because (by definition) $\rho\, (l/r_s)\, (r_s m_H) = c_L$, the continuum limit
value of $c_L$ is, within its error, the same as its finite $L$ value $12.61$. 
Finally the critical law of approach to the phase transition along an LCP can be
expressed as \cite{Irges:2009bi}
\be
\b = \b_c\, \frac{1}{1-k\,\left(\frac{\displaystyle c_L}
{\displaystyle \rho\,L}\right)^2}\, ,\label{crit}
\ee  
with $k$ some constant, confirming that it is indeed the quantity $\rho\, L$ which controls the limit.

To approach the continuum limit at a given value of $\g$
(at a sufficiently large $L$) it is convenient to first tune $\rho$ to an optimal value 
and then take $L$ as large as possible, the latter being the more time consuming step. 
The larger $\g$ is, that is the more isotropic the box becomes, the
smaller box is appropriate to sustain a string (in a large box a small
anisotropy is nearly unnoticeable), which means that a larger $\rho$ will be
needed. But the box can not be arbitrarily small. The string being a long
distance effect, will be unstable in a too small box. Thus, we expect to 
have an upper limit on $\rho$
\be
\rho < \rho_{\rm inst} \qquad \mbox{(at given $L$)} \,,
\ee 
beyond which some sign of instability should appear. Indeed, beyond a certain
$\rho$ the string tension turns negative, a typical sign of instability. 
On the other hand when $\g$ is small, the physical volume of the four
dimensional branes shrinks compared to the extra dimension and therefore a
large box, i.e. a small $\rho$ is
necessary to sustain a string. But the box can not be arbitrarily large
either. A lower bound 
\be
\rho > \rho_{\rm dr} \qquad \mbox{(at given $L$)} \label{lowrho}
\ee
on the possible values of $\rho$ exists, beyond which the
critical law \eq{crit} is not valid anymore and dimensional reduction is lost. 
The coefficients $\sigma$, $c_0$, $c_1$ plateau
at zero with only $c_2$ remaining non-trivial, as appropriate to a five dimensional
static potential. Also the ratio $F_5/F_4$ approaches a constant, which is 1
for the isotropic case.
 
The above discussion implies essentially that
only a box whose physical size and anisotropy
are within a certain allowed (correlated) range can describe a continuum four dimensional theory 
with a stable string. It turns out that 
in our case this range for the anisotropy parameter is around
\be
0.6 > \g > 0.5 \,, \label{bound1}
\ee
a rather narrow window. Our prime example, LCP A (for which $\g=0.55$), falls in the middle of this range. 
As either of the two bounds in \eq{bound1} is approached the system seems to require 
very quickly huge lattices in order that the continuum limit is reached.
For $\g=0.55$ the physical size of the lattice can be approximately in the range $2 > \rho > 0.5$.
Outside this range we observe either a negative string tension or five dimensional physics.  
Again, soon deviating from the LCP A value 
($\rho=0.625$), the continuum limit seems to demand larger lattices than we
can afford at the moment.
 
The remaining question is how to construct such a box that fits the
``universe''. The answer is in \fig{rhovar1} and \eq{crit}: 
we start with values $\rho_1$ and $L_1$ that describe four dimensional physics
(for LCP A this is realized at $\rho_1=0.625$ and $L_1=400$) and decrease
$\rho_2<\rho_1$ adjusting the lattice size so that $L_2\ge(\rho_1/\rho_2)L_1$.
Like this $\beta$ in \eq{crit} does not increase and the bound \eq{lowrho} is not violated. 
In particular these operations should make it possible to eventually reach
$\sqrt{\sigma}\, l > 1$.
Finally, it is interesting to observe that in this limit 
the scalar mass becomes much larger than the vector mass.

\section{Conclusions}
 
A five dimensional pure gauge theory on an anisotropic lattice can be
described by fluctuations around a mean-field background. The phase diagram
has a line of ultraviolet fixed points where the system can reduce to 
a collection of non-interacting Euclidean four-branes and the continuum limit
can be taken. The static potential together with the masses of
the lightest fields can be used to describe the system away from perturbation
theory. The four dimensional gauge coupling derived from the static potential
at short distance runs like an asymptotically free coupling while at long
distance it becomes the L\"uscher coefficient of a confining string. 
The four dimensional theory recovered in the continuum limit is not
a pure $SU(2)$ gauge theory. The potential shows confinement but also has a
large logarithmic contribution and the vector particle is lighter than the
scalar. This theory should correspond to a region in the parameter space of
the Georgi--Glashow model.

Even though we have computed all the observables analytically, we did the
continuum limit analysis numerically. Clearly, the success with which the model passed several
severe tests calls for further study, where the
lattice propagator is inverted analytically and the finite lattice sums are
performed explicitly. Then the continuum limit itself could be studied
analytically.

The static potential is computed using a MATLAB code.
Using an AMD Phenom II 3.2 GHz processor, the computational cost is
35 core-hours on a $L=200$ lattice and about $34$ core-days on a $L=400$ lattice.

{\bf Acknowledgments.} We thank R. Sommer for correspondence.
N. I. acknowledges the support of the Alexander von Humboldt
Foundation via a Fellowship for Experienced Researcher.

\bigskip

\begin{appendix}

\section{Appendix}

In this Appendix we summarize the expressions for the observables 
computed analytically in \cite{Irges:2009bi} and 
analyzed numerically here in the main text.

The propagator in momentum space is defined as
\be
{\tilde K}=-{{\tilde K}^{(hh)-1}}+{\tilde K}^{(vv)}
\ee
with
\bea
&& {\tilde K}^{(hh)} =  
-\delta_{p'p''}\delta_{\a'\a''}\frac{I_2({\overline h}_{0})}
{{\overline h}_{0} I_1({\overline h}_{0})} 
\left[1-\e\cdot\frac{{\overline h}_{0}}{I_2({\overline h}_{0})}
\left(\frac{I_2^2({\overline h}_{0})}
{I_1({\overline h}_{0})}-I_3({\overline h}_{0})\right)\right]\cdot 
{\rm diag}(1,1,1,1,0)\nonumber\\
&&  -\delta_{p'p''}\delta_{\a'\a''}
\frac{I_2({\overline h}_{05})}{{\overline h}_{05} I_1({\overline h}_{05})}
\left[1-\e\cdot\frac{{\overline h}_{05}}{I_2({\overline h}_{05})}\left(\frac{I_2^2({\overline h}_{05})}
{I_1({\overline h}_{05})}-I_3({\overline h}_{05})\right)\right]\cdot 
{\rm diag}(0,0,0,0,1)
\eea
where $\e=1$ for $\a'= 0$ and $\e=0$ for $\a'\ne 0$,
\bea
&&{\tilde K}^{(vv)}_{\a'\ne 0}=
\delta_{p'p''}\delta_{\a'\a''}(-2\frac{\b}{\g}{\ov}_0^2)\cdot \nonumber\\
&&\left(\begin{array} {cc} \sum^\prime c_{M'}-\frac{1}{\xi} s^2_{0/2} \hskip 1cm  2s_{0/2}s_{1/2}
 \hskip 1cm 2s_{0/2}s_{2/2} \hskip 1cm 2s_{0/2}s_{3/2}  \hskip 1cm 2s_{0/2}s_{5/2}
\g^2 \frac{\ov_{05}}{{\ov}_0} \\ 
2s_{1/2}s_{0/2}  \hskip 1cm \sum^\prime c_{M'}-\frac{1}{\xi}  s^2_{1/2}\hskip 1cm
2s_{1/2}s_{2/2} \hskip 1cm 2s_{1/2}s_{3/2} \hskip 1cm 2s_{1/2}s_{5/2}\g^2
\frac{\ov_{05}}{{\ov}_0} \\ 
2s_{2/2}s_{0/2} \hskip 1cm 2s_{2/2}s_{1/2} \hskip 1cm 
\sum^\prime c_{M'}-\frac{1}{\xi}  s^2_{2/2}\hskip 1cm
2s_{2/2}s_{3/2}   \hskip 1cm 2s_{2/2}s_{5/2}\g^2 \frac{\ov_{05}}{{\ov}_0}\\ 
2s_{3/2}s_{0/2}  \hskip 1cm 2s_{3/2}s_{1/2} \hskip 1cm
2s_{3/2}s_{2/2} \hskip 1cm \sum^\prime c_{M'}- \frac{1}{\xi}  s^2_{3/2}\hskip 1cm
2s_{3/2}s_{5/2}\g^2 \frac{\ov_{05}}{{\ov}_0} \\ 
2s_{5/2}s_{0/2} \g^2 \frac{\ov_{05}}{{\ov}_0} \hskip .5cm  2s_{5/2}s_{1/2} \g^2 \frac{\ov_{05}}{{\ov}_0}\hskip .5cm
2s_{5/2}s_{2/2} \g^2 \frac{\ov_{05}}{{\ov}_0}\hskip .5cm  2s_{5/2}s_{3/2}
\g^2 \frac{\ov_{05}}{{\ov}_0} \hskip .5cm
\g^2 \left(\sum^\prime c_{M'}- \frac{1}{\xi}  s^2_{5/2}\right)\\ 
\end{array} \right)\nonumber\\ \label{Kvva1}
\eea
and
\bea
&&{\tilde K}^{(vv)}_{\a' = 0}=
\delta_{p'p''}\delta_{\a'\a''}(-2\frac{\b}{\g}{\ov}_0^2)\cdot \nonumber\\
&&\left(\begin{array} {cc} \sum^\prime c_{M'} \hskip 1cm  2c_{0/2}c_{1/2}
 \hskip 1cm 2c_{0/2}c_{2/2} \hskip 1cm 2c_{0/2}c_{3/2}  \hskip 1cm 2c_{0/2}c_{5/2}
\g^2 \frac{\ov_{05}}{{\ov}_0} \\ 
2c_{1/2}c_{0/2}  \hskip 1cm \sum^\prime c_{M'} \hskip 1cm
2c_{1/2}c_{2/2} \hskip 1cm 2c_{1/2}c_{3/2} \hskip 1cm 2c_{1/2}c_{5/2}\g^2
\frac{\ov_{05}}{{\ov}_0} \\ 
2c_{2/2}c_{0/2} \hskip 1cm 2c_{2/2}c_{1/2} \hskip 1cm 
\sum^\prime c_{M'} \hskip 1cm
2c_{2/2}c_{3/2}   \hskip 1cm 2c_{2/2}c_{5/2}\g^2 \frac{\ov_{05}}{{\ov}_0}\\ 
2c_{3/2}c_{0/2}  \hskip 1cm 2c_{3/2}c_{1/2} \hskip 1cm
2c_{3/2}c_{2/2} \hskip 1cm \sum^\prime c_{M'} \hskip 1cm
2c_{3/2}c_{5/2}\g^2 \frac{\ov_{05}}{{\ov}_0} \\ 
2c_{5/2}c_{0/2} \g^2 \frac{\ov_{05}}{{\ov}_0} \hskip .5cm  2c_{5/2}c_{1/2} \g^2 \frac{\ov_{05}}{{\ov}_0}\hskip .5cm
2c_{5/2}c_{2/2} \g^2 \frac{\ov_{05}}{{\ov}_0}\hskip .5cm  2c_{5/2}c_{3/2}
\g^2 \frac{\ov_{05}}{{\ov}_0} \hskip .5cm
\g^2 \sum^\prime c_{M'}\\ 
\end{array} \right)\nonumber\\
\eea
The above propagator is written in the Lorentz gauge with parameter $\xi$.
Also, we have used the following notations and conventions:
The propagator is an object with the following index structure:
\be
{\tilde K} (p',M',\a'; p'',M'',\a'')
\ee
with $p$ the discrete momentum
\be
p_M = \frac{2\pi}{L} k_M \,,\quad k_M=0,1,\ldots, {L}-1 ,
\ee
$M=0,1,2,3,5$ a Euclidean index and $\a=0,1,2,3$ an index taking value in the Lie group $SU(2)$.
The Euclidean structure is  built in the matrix form. 
We use the notation $s_M = \sin{(p'_M)}$, $c_M = \cos{(p'_M)}$,  
$s_{M/2} = \sin{(p'_M/2)}$, $c_{M/2} = \cos{(p'_M/2)}$.
The only special case not explicitly shown in the matrices is that in the diagonal elements, 
$c_{5} = \g^2 \frac{{\ov}_{05}^2}{{\ov}_{0}^2} \cos{(p_5')} $.
On the diagonals $\sum'$ implies summation over all Euclidean indices leaving
out the one that corresponds to the row/column index of the term.
The functions $I_{1,2,3}$ are the usual Bessel functions of the type $I$.
The background values $\ov_0$ and $\ov_{05}$ 
along the four dimensional branes and extra dimension respectively, are determined by the extrimization of
\be
\frac{S_{\rm eff}[{\overline V},{\overline H}]}{L^5}=
-\frac{\b}{\g} \frac{(d-1)(d-2)}{2} \ov_0^4 - \b\g (d-1) \ov_0^2 \ov_{05}^2 + (d-1)
u({\overline h}_0) + u({\overline h}_{05}) + (d-1) {\overline h}_0 \ov_0 +{\overline h}_{05} \ov_{05}
\ee
which yields the conditions (primes denote derivatives with respect to the
argument)
\bea
&& \ov_0 = - u({\overline h}_0)'\,, \hskip 1cm 
{\overline h}_0=\frac{6\b}{\g}\ov_0^3+2\b\g \ov_0 \ov_{05}^2 \,,\nonumber\\
&& \ov_{05} = - u({\overline h}_{05})'\,, \hskip 1cm 
{\overline h}_{05}=8\b\g\ov_0^2 \ov_{05} \,,\label{MF_a}
\eea
for $d=5$. In the above we have introduced the function $u(x)=-\ln\left( \frac{2}{x} I_1(x) \right)$.
On an anisotropic lattice there are two inequivalent
Wilson loops of spatial length $r$. One along the four dimensional hyperplanes for which the
static potential is given by
\bea
V_4(r)  &=& -2\log(\ov_0)-\frac{1}{2\ov_0^2}\frac{1}{L^4}
\times\left\{\sum_{p_{M\ne 0}',p_0'=0}\left[\frac{1}{3}\sum_k(2\cos(p_k'r)+2)\right] 
C^{-1}_{00}(p',0) \right. \nonumber\\
&&\left. + 3\sum_{p_{M \ne 0}',p_0'=0}\left[\frac{1}{3}\sum_k(2\cos(p_k'r)-2)\right]
\frac{1}{C_{00}(p',1)}\right\} \,.\label{V4}
\eea
The one along the extra dimension is given by
\bea
V_5(r) &=& -2\log(\ov_0)\nonumber\\
&-&\frac{1}{2\ov_0^2}\frac{1}{L^4}
\sum_{p_{M\ne 0}',p_0'=0}\left\{\left[2\cos(p_5'r)+2\right] 
C^{-1}_{00}(p',0) + 3\left[2\cos(p_5'r)-2\right]
\frac{1}{C_{00}(p',1)}\right\} \,. \label{V5}\nonumber\\
\eea
The matrix $C$ in \eq{V4} and \eq{V5} is defined from
\be
\tilde{K}(p',M',\a';p'',M'',\a'') = \delta_{p'p''}\delta_{\a'\a''}
C_{M'M''}(p',\a').
\ee
The scalar mass is derived from the correlator
\be
C_H(t) = \frac{1}{L^5}(P_0^{(0)})^2\sum_{p_0'}\cos{(p_0't)}
\sum_{p_5'}|{\tilde \D}^{(L)}(p_5')|^2{\tilde K}^{-1}
\Bigl((p_0',{\vec 0},p_5'),5,0; (p_0',{\vec 0},p_5'), 5,0\Bigr)
\,, \label{H}
\ee
where $P_0^{(0)}$ is the associated Polyakov loop evaluated on the background 
(being an overall constant, its value is irrelevant for the mass) and
\be
{\tilde \D}^{(L)}(p) = \sum_{m=0}^{L-1}
\frac{e^{ip(m+1/2)}}{\ov_{05}} \,.
\ee
For the $W$ gauge boson mass we define the contractions 
\be
{\overline K}^{-1}((p_0',{\vec p}'),5,\a)=
\sum_{p_5',p_5''}{\tilde \D}^{(L)}(p_5'){\tilde \D}^{(L)}(-p_5'') K^{-1}(p'',5,\a; p',5,\a)\nonumber\\
\ee
and
\be
{\overline {\overline K}}^{-1}(t,{\vec p}',\a) = \sum_{p_0'}e^{ip_0't}
{\overline K}^{-1}((p_0',{\vec p}'),5,\a) .
\ee
in terms of which the vector correlator reads
 \be
C_{V}(t)=
\frac{768}{L^{10}} (P_0^{(0)})^4 (\ov_0(0))^4 
\sum_{{\vec p}'} \sum_k \sin^2(p_k')
\left({\overline {\overline K}}^{-1}(t,{\vec p}',1)\right)^2 \,.\label{W}
\ee
The lightest state's masses are read off to second order from
the exponential decay of the correlators according to
\be
m \simeq \lim_{t\to \infty} \ln \frac{C^{(1)} (t)+C^{(2)} (t)}{C^{(1)} 
(t-1)+C^{(2)} (t-1)} \,.\label{mcor2}
\ee
The scalar mass is non-trivial at first order while the vector mass is non-trivial at second order.

\section{Appendix}

Here we write the discretization formulae for the computation of the local
coefficients of \eq{Short} and \eq{Long} to fit the potential along the four dimensional hyperplanes.
We introduce the dimensionless potential $a_4V(r)=\oV(x)$, define $x=r/a_4$ and the
dimensionless coefficients $\overline{\sigma}=a_4^2\sigma$,
$\overline{c_2}=c_2/a_4$ and $\overline{c_0}=a_4c_0$. 
$c_1$ is already dimensionless.

We use derivatives of the potential improved to O$(a_4^2)$:
\bea
\oV^\prime(x) & = & \{\oV(x+1)-\oV(x-1)\}/2 \,, \label{dV} \\
\oV^{\prime\prime}(x) & = & \oV(x+1)+\oV(x-1)-2\oV(x) \,, \label{d2V} \\
\oV^{\prime\prime\prime}(x) & = & \{\oV(x+2)-\oV(x-2)-2[\oV(x+1)-\oV(x-1)]\}/2 \,,
\label{d3V} \\
\oV^{\prime\prime\prime\prime}(x) & = & \{\oV(x+2)+\oV(x-2) -
4[\oV(x+1)+\oV(x-1)] + 6\oV(x) \} \,. \label{d4V}
\eea
In terms of these derivatives the coefficients in \eq{Short} are estimated to
be
\bea
c_1(x) & = & 2x^3\oV^{\prime\prime}(x) + 1/2x^4\oV^{\prime\prime\prime}(x) \,,
\nonumber\\
\overline{c_2}(x) & = & - 1/2x^4\oV^{\prime\prime}(x) -
1/6x^5\oV^{\prime\prime\prime}(x) \,,
\eea
and those of \eq{Long}
\bea
\overline{\sigma}(x) & = & \oV^\prime(x) + 3x\oV^{\prime\prime}(x) +
3/2x^2\oV^{\prime\prime\prime}(x) + 1/6x^3\oV^{\prime\prime\prime\prime}(x)
\,, \\
{\overline{c_0}}(x) & = & -6x^2\oV^{\prime\prime}(x) -
4x^3\oV^{\prime\prime\prime}(x) - 1/2x^4\oV^{\prime\prime\prime\prime}(x)\, ,\\
c_1(x) & = & -4x^3\oV^{\prime\prime}(x) - 7/2x^4\oV^{\prime\prime\prime}(x) -
1/2x^5\oV^{\prime\prime\prime\prime}(x) \,, \\
\overline{c_2}(x) & = & 1/2x^4\oV^{\prime\prime}(x) +
1/2x^5\oV^{\prime\prime\prime}(x) + 1/12x^6\oV^{\prime\prime\prime\prime}(x)\,.
\eea

\end{appendix}


\end{document}